\documentclass[aps,prc,superscriptaddress,twoside,twocolumn,showpacs]{revtex4}
\usepackage{amsmath,amssymb}
\usepackage{graphicx}
\usepackage{graphics}
\usepackage{epsfig}
\begin{document}

\title{Comparison of inclusive $\boldsymbol{K^+}$ production in
proton-proton and proton-neutron collisions}

\author{Yu.~Valdau}\email{y.valdau@fz-juelich.de}
\affiliation{High Energy Physics Department, Petersburg Nuclear
Physics Institute, RU-188350 Gatchina, Russia} \affiliation{Institut
f\"ur Kernphysik and J\"ulich Centre for Hadron Physics,
Forschungszentrum J\"ulich, D-52425 J\"ulich, Germany}
\author{V.~Koptev}
\affiliation{High Energy Physics Department, Petersburg Nuclear
Physics Institute, RU-188350 Gatchina, Russia}
\author{S.~Barsov}
\affiliation{High Energy Physics Department, Petersburg Nuclear
Physics Institute, RU-188350 Gatchina, Russia}
\author{M.~B\"uscher}
\affiliation{Institut f\"ur Kernphysik and J\"ulich Centre for Hadron
Physics, Forschungszentrum J\"ulich, D-52425 J\"ulich, Germany}
\author{A.~Dzyuba}
\affiliation{High Energy Physics Department, Petersburg Nuclear
Physics Institute, RU-188350 Gatchina, Russia}
\author{M.~Hartmann}
\affiliation{Institut f\"ur Kernphysik and J\"ulich Centre for Hadron
Physics, Forschungszentrum J\"ulich, D-52425 J\"ulich, Germany}
\author{A.~Kacharava}
\affiliation{Institut f\"ur Kernphysik and J\"ulich Centre for Hadron
Physics, Forschungszentrum J\"ulich, D-52425 J\"ulich, Germany}
\author{I.~Keshelashvili}
\affiliation{Department of Physics, University of Basel,
Klingelbergstrasse 82, CH-4056, Basel, Switzerland}
\author{A.~Khoukaz}
\affiliation{Institut f\"ur Kernphysik, Universit\"at M\"unster,
D-48149 M\"unster, Germany}
\author{S.~Mikirtychiants}
\affiliation{High Energy Physics Department, Petersburg Nuclear
Physics Institute, RU-188350 Gatchina, Russia} \affiliation{Institut
f\"ur Kernphysik and J\"ulich Centre for Hadron Physics,
Forschungszentrum J\"ulich, D-52425 J\"ulich, Germany}
\author{M.~Nekipelov}
\affiliation{Institut f\"ur Kernphysik and J\"ulich Centre for Hadron
Physics, Forschungszentrum J\"ulich, D-52425 J\"ulich, Germany}
\author{A.~Polyanskiy}
\affiliation{Institut f\"ur Kernphysik and J\"ulich Centre for Hadron
Physics, Forschungszentrum J\"ulich, D-52425 J\"ulich, Germany}
\affiliation{Institute for Theoretical and Experimental Physics,
RU-117218 Moscow, Russia}
\author{F.~Rathmann}
\affiliation{Institut f\"ur Kernphysik and J\"ulich Centre for Hadron
Physics, Forschungszentrum J\"ulich, D-52425 J\"ulich, Germany}
\author{H.~Str\"oher}
\affiliation{Institut f\"ur Kernphysik and J\"ulich Centre for
Hadron Physics, Forschungszentrum J\"ulich, D-52425 J\"ulich,
Germany}
\author{Yu.~N.~Uzikov}
\affiliation{Laboratory of Nuclear Problems, JINR, RU-141980 Dubna,
Russia}
\author{C.~Wilkin}\email{cw@hep.ucl.ac.uk}
\affiliation{Physics and Astronomy Department, UCL, London, WC1E 6BT,
United Kingdom}
\date{\today}

\begin{abstract}
The momentum spectra of $K^+$ produced at small angles in
proton-proton and proton-deuteron collisions have been measured at
four beam energies, 1.826, 1.920, 2.020, and 2.650~GeV, using the
ANKE spectrometer at COSY-J\"ulich. After making corrections for
Fermi motion and shadowing, the data indicate that $K^+$ production
near threshold is stronger in $pp$- than in $pn$-induced reactions.
However, most of this difference could be made up by the unobserved
$K^0$ production in the $pn$ case.
\end{abstract}

\pacs{13.75.-n, 
      14.40.Df, 
      25.40.Ve} 

\maketitle

%
%
\section{Introduction}
\label{introduction}

A large series of results on $K^+$ production in
proton-nucleus~\cite{BUS2004} and nucleus-nucleus
collisions~\cite{LOP2007} has been compiled in the low energy region.
Although different nuclear models have been used to describe these
data~\cite{RUD2005, GAU2010}, the crucial ingredients in any
theoretical interpretation must be the cross sections for strangeness
production in proton-proton and proton-neutron reactions.
Parameterizations of the existing experimental data, or calculations
within a meson-exchange model~\cite{TSU2000}, have generally been
employed for this purpose~\cite{HAR2005}.

\begin{figure}[hbt]\centering
\includegraphics[width=.99\columnwidth,clip=]{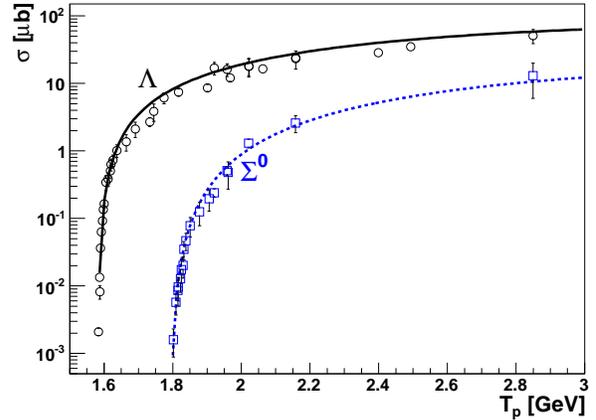}
\caption{\label{fig:totcr}(Color online) Compilation of total cross
sections for the $pp\to K^+p\Lambda$ (black circles), and $pp\to
K^+p\Sigma^0$ (blue squares) reactions. The experimental data are
taken from Refs.~\cite{BAL1996,BIL1998,SEW1999,KOW2004,
ABD2006,VAL2007,LOU1961,VAL2010,BAL1988}. Parametrizations of
experimental data for $\Lambda$ and $\Sigma^0$ production from
Ref.~\cite{SIB2006} are presented by solid black and dashed blue
lines, respectively. }
\end{figure}

There are three strangeness-conserving $K^+$ production channels that
are open in $pp$ collisions close to threshold. The one that has been
most widely investigated is $pp\to K^+p\Lambda$, for which data on
the total cross section, Dalitz plots, and angular spectra have been
published~\cite{BAL1996,BIL1998,SEW1999,KOW2004,ABD2006,VAL2007,LOU1961,VAL2010,BAL1988,ABD2010a}.
In the $pp\to K^+p \Sigma^0$ case, values of the total cross
sections~\cite{SEW1999,KOW2004} and angular
distributions~\cite{ABD2010a} are available.
Parameterizations~\cite{SIB2006} of the energy dependence of the
total cross sections for $\Lambda$ and $\Sigma^0$ production are
shown in Fig.~\ref{fig:totcr} alongside the available experimental
data. Measurements also exist of the $pp\to K^+n \Sigma^+$ total
cross section~\cite{ROZ2006, VAL2007, VAL2010} and studies of the
$pp\to K^0p \Sigma^+$ channel have been
undertaken~\cite{ABD2004,ABD2007}.
%
%
%
%

In contrast, very little is known experimentally about $K^+$
production in proton-neutron collisions. There are a few $np\to K^+p
\Sigma^-$ data points obtained with a neutron beam~\cite{ANS1973},
though only at 3.4~GeV and above. There seem to be no measurements of
the $pn \to K^+n \Lambda$ or $pn \to K^+n \Sigma^0$ reaction channels
recorded in the literature. However, some data on $pn \to K^+n
\Sigma^-$ have been obtained using the COSY-ANKE
spectrometer~\cite{Sigmaminus} and results on the $pn \to K^+n
\Lambda$ reaction close to threshold should eventually be available
from this facility~\cite{DZY2010} as will $pn\to K^0 p \Lambda$ from
the COSY-TOF spectrometer~\cite{KRA2009}.

Several attempts have been made to deduce the ratio of $K^+$
production in $pp$ and $pn$ collisions,
$\sigma_{pn}^{K^+}/\sigma_{pp}^{K^+}$, by analyzing data obtained
using different nuclear targets and beams. The rather inconclusive
value of $\sigma_{pn}^{K^+}/\sigma_{pp}^{K^+}=5 \pm 7.5$ was obtained
with a 3~GeV proton beam incident on a range of
nuclei~\cite{BER1958}. The comparison of the $K^+$ production rate on
a NaF target with proton and deuteron beams of energy 2.1~GeV per
nucleon yielded a ratio $\sigma(d\,\textrm{NaF} \to
KX)/\sigma(p\,\textrm{NaF} \to KX) =1.3\pm0.2$, from which it was
concluded that
$\sigma_{pn}^{K^+}/\sigma_{pp}^{K^+}<1$~\cite{SCH1989}. The ratio of
double-differential $K^+$ production cross sections measured with
carbon and hydrogen targets at 2.5~GeV was interpreted as evidence
that $\sigma_{pn}^{K^+}/\sigma_{pp}^{K^+} \approx 1$~\cite{DEB1996}.
However, an analysis of the double-differential cross section for
$K^+$ production with a 2.02~GeV proton beam incident on a deuterium
target gave the much larger but very model-dependent estimate of
$\sigma_{pn}^{K^+}/\sigma_{pp}^{K^+} \approx 3-4$~\cite{BUS2004}.

In the absence of reliable experimental proton-neutron data, it is
interesting to see if theory can offer any guidance. The models that
have been used are mainly of the one-meson-exchange variety, though
these are very uncertain even for the $pp\to K^+p \Lambda$ reaction
since there is no consensus as to which exchanges need to be
included. The results of the COSY-TOF collaboration, and in
particular the evidence for the excitation of $N^*$ isobars, suggest
the dominance of non-strange meson exchange~\cite{ABD2010b}. On the
other hand, measurements of the spin-transfer parameter $D_{NN}$ have
been interpreted as indicating that strange meson exchange is the
more important~\cite{BAL1999}. The picture is even less clear in the
$pn$ case, with theoretical estimates of $\sigma(pn\to K^+n
\Lambda)/\sigma(pp\to K^+p \Lambda)$ ranging from 0.25 to
5~\cite{FAL2005}, or around 2~\cite{TSU2000} or 3~\cite{IVA2006},
depending upon the assumptions made.

The situation can only be clarified by further experimental work.
This is reported here in the form of inclusive $K^+$ momentum spectra
measured on hydrogen and deuterium targets at four proton beam
energies. The comparison of the results on the two targets allows one
to make estimates of $\sigma_{pn}^{K^+}/\sigma_{pp}^{K^+}$. The
experimental setup used to carry out the measurements is described in
Sec.~\ref{experiment}. The results for the two targets are presented
in Sec.~\ref{results}, where it is clearly seen that the small angle
production rates from deuterium fall well below a factor of two times
those on hydrogen.

The $pp$ data are modeled in Sec.~\ref{model} in terms of
contributions from the $pp\to K^+p\Lambda$ and $pp\to K^+N\Sigma$
channels, with the normalizations chosen to fit the known total cross
sections. Such an approach is very successful at the three lower
energies but fails badly at 2.65~GeV when heavier hyperons and extra
pions can be produced. In order to clarify the effects of the
kinematics, the cross sections were smeared over the Fermi motion in
the deuteron, though the resultant changes were fairly minor. Taking
into account a small shadowing correction, it is seen in
Sec.~\ref{discussion} that $K^+n \Lambda$ production on the neutron
is about half of that for $K^+p \Lambda$ on the proton. However,
since half the strength in the $pn$ case should emerge in the
$K^0 p \Lambda$ channel, our result is consistent with strangeness
production near threshold being of similar size for both isospin
$I=1$ and $I=0$. Our conclusions and outlook for further work are
reported in Sec.~\ref{conclusions}.
%
%

\section{Experimental method}
\label{experiment}

The experiments were carried out at the Cooler-Synchrotron
COSY-J\"{u}lich~\cite{MAI97} using unpolarized proton beams incident
on hydrogen and deuterium cluster-jet targets~\cite{KHO99}. The
resulting reaction products were detected in the ANKE magnetic
spectrometer~\cite{BAR01}, which is situated in a chicane inside the
COSY ring. The spectrometer uses three dipole magnets. \emph{D1} and
\emph{D3} divert the circulating beam onto the ANKE target and back
into the COSY ring, respectively, while \emph{D2} is the analyzing
magnet. All data with the hydrogen target and those with deuterium at
$T_p=2.650$~GeV were collected using the same magnetic field in
\emph{D2} ($B_{\text{max}}=1.568$~T). Deuterium data at $1.826$,
$1.920$ and $2.020$~GeV were taken using $B_{\text{max}}=1.445$,
$1.505$ and $1.570$~T, respectively.

\begin{figure}[hbt]
\begin{center}
\includegraphics[scale=0.55]{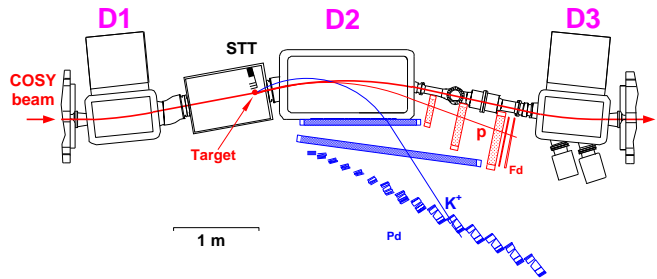}
\end{center}
\caption{(Color online) Sketch of the relevant parts of the ANKE
detector system, showing the positions of the two bending magnets
\emph{D1} and \emph{D3} and the target placed before the analyzing
magnet \emph{D2}. Only information obtained from the forward (Fd),
positive side (Pd) detectors, and the Silicon Tracking Telescope
(STT) was used in this experiment. Typical trajectories, leading to
the measurement of protons and $K^+$, are also
shown.\label{fig:anke}}
\end{figure}

The layout of the relevant elements of the ANKE detector systems is
sketched in Fig.~\ref{fig:anke}. Only information provided by the
positive detector (Pd), forward detector (Fd) and a prototype of the
silicon tracking telescope (STT) was used in the present analysis.
The Pd detector system consists of start and stop counters for
time-of-flight measurements, two multiwire proportional chambers for
momentum reconstruction and background suppression, and range
telescopes for $K^+$ identification.

%
%
%
%

Fifteen range telescopes were placed in the focal plane of the
\emph{D2} magnet. Each of them consists of a stop counter, two copper
degraders, a counter for measuring the energy loss ($\Delta E$
counter) and one for detecting the $K^+$ decay products (veto
counter). The thickness of the first degrader was chosen such that
the kaon deposits maximal energy in the $\Delta E$ counter and stops
either at its edge or in the second degrader. Measurements of the
time difference between the $K^+$ in the stop counter and its decay
products in the veto counter provide a clear $K^+$ identification
even if the background from pions and protons is $10^6$ times
higher~\cite{BUS02}. The efficiency of each range telescope, which
was between 10\% and 30\% depending upon the particular telescope,
the parameters of the cuts, and the trigger conditions, were
determined from the experimental data. The overall uncertainty from
telescope to telescope arising from this efficiency correction was
estimated with an accuracy of $\approx 15$\% for the $pd$ data and
the $pp$ at $2.65$~GeV, and $\approx 10$\% for the other $pp$ runs.
Full details of $K^+$ identification using the ANKE range telescopes
can be found in Ref.~\cite{BUS02}.

The cross section normalizations were established from the
information supplied by the Fd detector system, as described in
Ref.~\cite{DYM04}. The Fd consists of three multiwire chambers and a
hodoscope of scintillators. In the hydrogen-target experiments, and
for deuterium at $2.650$~GeV, the first multiwire proportional
chamber was replaced by a drift chamber. For the hydrogen
experiments, a dedicated prescaled trigger was used to monitor the
$pp$ elastic scattering rate throughout the experimental runs.
Elastic scattering events were identified in the angular range
$6.0^{\circ}$--$\,9.0^{\circ}$ in the forward detector using the
missing-mass technique. The luminosity was then evaluated on the
basis of the known $pp$ differential cross sections with an overall
normalization error of 7\%~\cite{VAL2010}.

A similar method was employed to normalize the 2.650~GeV $pd$ data.
The prescaled rate for the production of fast protons from
proton-deuteron collisions was continuously monitored in the forward
detector over the angular range $7.0^{\circ}$--$\,9.0^{\circ}$.
Because it was not possible to distinguish between elastic scattering
and deuteron break-up events, the experimental count rate was
converted into luminosity estimates through calculations carried out
within the Glauber-Sitenko theory~\cite{UZI01}. The associated
overall normalization error should not exceed 15\%~\cite{YAS02}.

The coincidence of signals from the silicon tracking telescope and
forward detector was used to measure $pd$ elastic scattering at
$1.826, 1.920$ and $2.020$~GeV. The STT consists of three silicon
detectors placed in ultra-high vacuum, with the first layer being
5~cm from the beam-target interaction point~\cite{LEH04}. The
identification of the deuteron in the STT, together with a proton in
the Fd, allows one to select unambiguously elastic $pd\to pd$ events.
However, uncertainties in the evaluation of the acceptance
\emph{etc.}\ means that the overall absolute normalization error was
$20$\%, though the relative error between different beam energies was
at most $5$\%~\cite{BAR04}.

%
%

\section{Experimental Results}
\label{results}

The double-differential cross section for $K^+$ production in
each momentum bin $\Delta p$ is evaluated from
\begin{equation}
d\sigma^{K^+} \equiv \frac{d^{\,2}\sigma_{K^+}}{d\Omega\,dp}(T_p)=\frac{N_{K^{+}}}{\Delta{p}\,\Delta\Omega}
\frac{1}{L^{\text{tot}}\,\epsilon_{K^{+}}}, \label{eq:d2sigmadOmdP}
\end{equation}
where $N_{K^{+}}$ is the number of $K^+$ detected in a solid angle
$\Delta\Omega$ and $L^{\text{tot}}$ is the integrated luminosity. The
efficiency of $K^+$ identification, $\epsilon_{K^{+}}$, is estimated
from
\begin{equation}
\epsilon_{K^+} =
\epsilon^{\text{tel}}\times\epsilon^{\text{scint}}\times
\epsilon^{\text{MWPC}}\times\epsilon^{\text{acc}}\,.
 \label{eq:effK}
\end{equation}
The scintillator ($\epsilon^{\text{scint}}$) and MWPC
($\epsilon^{\text{MWPC}}$) efficiencies are determined from the
experimental data. The kaon acceptance, including corrections for
decay in flight, ($\epsilon^{\text{acc}}$) is determined using
simulations. The range-telescope efficiency $\epsilon^{\text{tel}}$,
whose value is extracted from calibration data on $K^+p$
coincidences, represents the main systematic uncertainty in the cross
section evaluation~\cite{VAL2010}.

\begin{figure}[hbt]\centering
\includegraphics[width=.99\columnwidth,clip=]{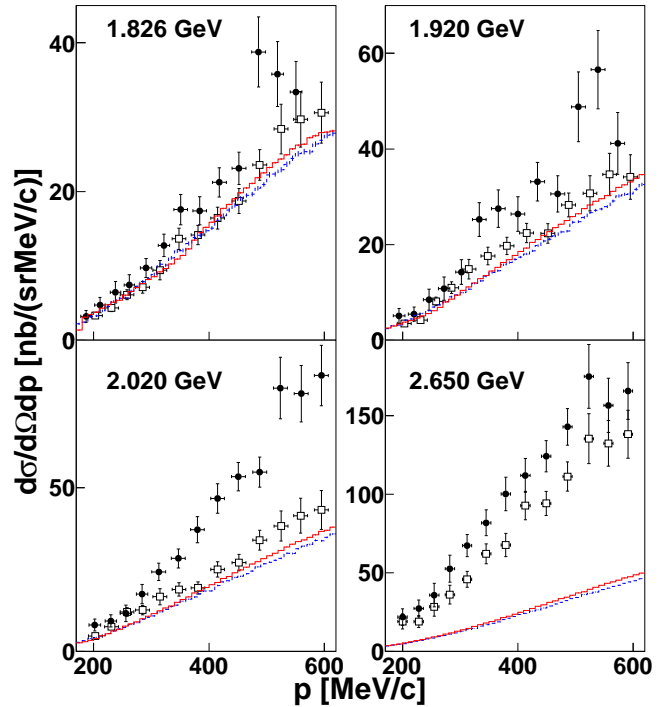}
\caption{\label{fig:dcr} (Color online) Momentum spectra of $K^+$
produced in $pp$ (open circles) and $pd$ (closed circles) collisions
at 1.826, 1.920, 2.020, and 2.650~GeV. Only statistical errors are
shown. The red lines represent the description of the $pp$ data
within the three-channel model whereas for the blue dashed ones the
predictions have been smeared over the deuteron Fermi momentum and a
correction made for the shadowing in the deuteron.}
\end{figure}

The laboratory double-differential cross section for $K^+$ production
on hydrogen (open symbols) and deuterium (closed symbols) are
presented at the four different beam energies in Fig.~\ref{fig:dcr}
as functions of the kaon momentum. The cross sections represent
averages over $K^+$ laboratory production angles up to $4^{\circ}$;
the measured values are also collected in Table~\ref{tab:TDDCr}.

%
%
%
%

The double-differential cross sections for $K^+$ production in $pd$
and $pp$ collisions, shown in Fig.~\ref{fig:dcr}, have rather similar
shapes and their ratio depends only weakly on the kaon momentum.
Because the $pd$ data at the three lower energies were taken with
different values of the \emph{D2} magnetic field from those for $pp$,
the associated momentum bins are slightly different. In these cases
the ratio of the $K^+$ production cross sections was evaluated by
assuming a linear variation from one bin to the next. However,
corrections for the data measured at 2.020~GeV are minimal.
%
%
\begin{table*}[ht]
\begin{center}
\begin{tabular}{|c|c|c|c|c|c|c|c|c|c|}
\hline
\multicolumn{2}{|c|}{$T_p=1.826$~GeV}&
\multicolumn{2}{c|}{$T_p=1.920$~GeV}&
\multicolumn{2}{c|}{$T_p=2.020$~GeV}&
\multicolumn{4}{c|}{$T_p=2.650$~GeV}\\
\hline
$p$   & $d^2\sigma_{pd}^{K^+}\!\!/ d\Omega dp$ &
$p$   & $d^2\sigma_{pd}^{K^+}\!\!/ d\Omega dp$ &
$p$   & $d^2\sigma_{pd}^{K^+}\!\!/ d\Omega dp$ &
$p$   & $d^2\sigma_{pd}^{K^+}\!\!/ d\Omega dp$ &
$d^2\sigma_{pp}^{K^+}\!\!/ d\Omega dp$ & $d\sigma_{pd}^{K^+}/d\sigma_{pp}^{K^+}$\\
MeV/$c$ & nb/(sr\,MeV/$c$)      &
MeV/$c$ & nb/(sr\,MeV/$c$)      &
MeV/$c$ & nb/(sr\,MeV/$c$)      &
MeV/$c$ & \multicolumn{2}{c|}{nb/(sr\,MeV/$c$)}&
\\
\hline
187 $\pm$ 9 & 3.2 $\pm$ 0.8 & 194 $\pm$ 10 & 5.1 $\pm$ 1.6 & 202 $\pm$ 10 & 8.1 $\pm$ 1.8 & 200 $\pm$ 8 & 21.9 $\pm$ 5.3 & 19.1 $\pm$ 4.7 & 1.14 $\pm$ 0.21\\
211 $\pm$ 9 & 4.7 $\pm$ 1.0 & 220 $\pm$ 10 & 5.5 $\pm$ 1.5 & 230 $\pm$ 10 & 9.3 $\pm$ 1.8 & 228 $\pm$ 8 & 27.4 $\pm$ 5.3 & 19.0 $\pm$ 3.8 & 1.34 $\pm$ 0.22\\
238 $\pm$ 9 & 6.5 $\pm$ 1.4 & 246 $\pm$ 10 & 8.5 $\pm$ 2.2 & 256 $\pm$ 10 & 11.6 $\pm$ 2.5 & 255 $\pm$ 8 & 35.9 $\pm$ 7.5 & 28.4 $\pm$ 6.0 & 1.38 $\pm$ 0.12\\
262 $\pm$ 9 & 7.4 $\pm$ 1.4 & 272 $\pm$ 10 & 10.8 $\pm$ 2.4 & 284 $\pm$ 10 & 17.5 $\pm$ 2.9 & 282 $\pm$ 8 & 52.6 $\pm$ 8.6 & 36.1 $\pm$ 6.0 & 1.39 $\pm$ 0.11\\
291 $\pm$ 10 & 9.7 $\pm$ 1.3 & 302 $\pm$ 10 & 14.3 $\pm$ 2.6 & 313 $\pm$ 12 & 24.3 $\pm$ 2.6 & 312 $\pm$ 8 & 67.4 $\pm$ 7.1 & 45.9 $\pm$ 5.0 & 1.48 $\pm$ 0.11\\
322 $\pm$ 10 & 12.8 $\pm$ 1.5 & 333 $\pm$ 12 & 25.3 $\pm$ 3.5 & 347 $\pm$ 12 & 28.4 $\pm$ 2.9 & 345 $\pm$ 8 & 81.8 $\pm$ 8.2 & 62.2 $\pm$ 6.3 & 1.35 $\pm$ 0.09\\
351 $\pm$ 12 & 17.6 $\pm$ 2.0 & 366 $\pm$ 12 & 27.5 $\pm$ 3.9 & 380 $\pm$ 12 & 37.2 $\pm$ 4.0 & 379 $\pm$ 8 & 100.2 $\pm$ 10.6 & 67.7 $\pm$ 7.3 & 1.52 $\pm$ 0.09\\
384 $\pm$ 12 & 17.4 $\pm$ 1.9 & 400 $\pm$ 12 & 26.4 $\pm$ 3.6 & 415 $\pm$ 12 & 46.7 $\pm$ 4.5 & 413 $\pm$ 8 & 112.0 $\pm$ 10.8 & 92.8 $\pm$ 9.0 & 1.32 $\pm$ 0.07\\
418 $\pm$ 12 & 21.2 $\pm$ 2.0 & 434 $\pm$ 12 & 33.2 $\pm$ 4.0 & 451 $\pm$ 12 & 53.4 $\pm$ 4.3 & 449 $\pm$ 8 & 124.2 $\pm$ 10.0 & 94.2 $\pm$ 7.7 & 1.40 $\pm$ 0.08\\
452 $\pm$ 12 & 23.1 $\pm$ 2.2 & 469 $\pm$ 12 & 30.6 $\pm$ 3.8 & 488 $\pm$ 12 & 54.7 $\pm$ 4.6 & 486 $\pm$ 8 & 142.9 $\pm$ 11.7 & 111.3 $\pm$ 9.2\phantom{1} & 1.29 $\pm$ 0.06\\
486 $\pm$ 12 & 38.8 $\pm$ 4.7 & 505 $\pm$ 12 & 48.8 $\pm$ 7.3 & 524 $\pm$ 12 & 80.4 $\pm$ 9.4 & 523 $\pm$ 8 & 175.0 $\pm$ 20.4 & 135.4 $\pm$ 15.9 & 1.31 $\pm$ 0.08\\
519 $\pm$ 10 & 35.8 $\pm$ 4.4 & 539 $\pm$ 12 & 56.6 $\pm$ 8.2 & 560 $\pm$ 12 & 78.7 $\pm$ 8.6 & 557 $\pm$ 8 & 156.6 $\pm$ 17.2 & 132.4 $\pm$ 14.6 & 1.29 $\pm$ 0.11\\
551 $\pm$ 10 & 33.4 $\pm$ 4.1 & 573 $\pm$ 12 & 41.2 $\pm$ 6.5 & 595 $\pm$ 12 & 84.2 $\pm$ 9.2 & 591 $\pm$ 8 & 165.7 $\pm$ 18.1 & 138.3 $\pm$ 15.2 & 1.21 $\pm$ 0.08\\

\hline

\end{tabular}
\end{center}
\caption{The double-differential cross sections for $K^+$ production
measured in $pd$ collisions over the interval $\vartheta<4^{\circ}$
as a function of the kaon momentum $p$ at four beam energies. At
$2.650$~GeV the differential cross sections on hydrogen are also
presented, as are the $pd/pp$ ratios. The errors do not include the
overall systematic uncertainty associated with the
normalizations.\label{tab:TDDCr}}
\end{table*}
%
%

Since the $pp$ and $pd$ data at $2.650$~GeV were collected under
identical conditions, many of the factors in
Eq.~\eqref{eq:d2sigmadOmdP} cancel out and the ratio of the cross
sections for $K^+$ production on the deuteron and proton may then be
written as
\begin{equation}
d\sigma^{K^+}_{pd}/d\sigma^{K^+}_{pp}
=(N_{pd}^{K^+}/N_{pp}^{K^+})\times\,(L_{pp}^{\text{tot}}/L_{pd}^{\text{tot}})\,.
\end{equation}
Numerical values for the ratio at this energy are also presented in
Table~\ref{tab:TDDCr}.

\begin{table}[ht]
\begin{center}
\begin{tabular}{|c|c|c|c|}
\hline
$T_p$ GeV & $d\sigma_{pd}^{K^+}\!/ d\sigma_{pp}^{K^+}$ & $\Delta_{\text{aver}}$ & $\Delta_{\text{syst}}$\\
\hline
1.826     & 1.28                  &  0.03          & 0.28\\
1.920     & 1.38                  &  0.06          & 0.30\\
2.020     & 1.65                  &  0.10          & 0.36\\
2.650     & 1.34                  &  0.04          & 0.23\\
\hline
\end{tabular}
\end{center}
\caption{The mean ratio of the double-differential cross sections for
$K^+$ production in $pd$ and $pp$ collisions at four beam energies.
The errors arising from averaging over the kaon momentum
($\Delta_{\text{aver}}$) and the normalization uncertainty
($\Delta_{\text{norm}}$) are also presented. The relative
normalization uncertainty between the ratios measured at $1.826$,
$1.920$, and $2.020$~GeV is estimated to be 9\%. \label{tab:TRatio}}
\end{table}

The momentum dependence of the ratio
$d\sigma^{K^+}_{pd}/d\sigma^{K^+}_{pp}$ is relatively weak at all
four energies. Therefore, for comparison to the model discussed in
Sec.~\ref{model}, we calculate the weighted average of the ratio of
the differential cross sections over the $K^+$ momentum range. The
numerical values of the ratio are given in Table~\ref{tab:TRatio} and
presented graphically in Fig.~\ref{fig:money}. The averaging error
$\Delta_{\text{aver}}$ given in the table takes into account, not
only the errors for individual points, but also the momentum
dependence of the ratio.

%
%

\section{Three-channel model}
\label{model}

There is very little information on the production of $K^+$ in
association with excited hyperons and almost nothing on reactions
where an extra pion is produced. As a consequence, we model the data
purely in terms of $K^+p \Lambda$, $K^+p \Sigma^0$, and $K^+n
\Sigma^+$ final states, though this will clearly be insufficient at
the highest energy. Kaon production on a proton is assumed to be the
sum of contributions from these three channels. In the estimation of
the $\Lambda$ contribution, the phase space was modified by the
$p\Lambda$ final state interaction and the $N^*(1650)$ resonance, as
suggested in Ref.~\cite{ABD2006}. On the other hand, simple phase
space descriptions were used for the two $\Sigma$ channels. The
predictions were normalized to the parameterizations of the total
cross sections for $\Lambda$ and $\Sigma$ production~\cite{SIB2006}.
For this purpose, the total cross section for $\Sigma^+$ production
was  assumed to be a factor $0.7$ smaller than that for  $\Sigma^0$
at the same excess energy~\cite{VAL2010}.

The $pp$ calculations describe reasonably well the experimental
$K^+$ momentum spectra at the three lower energies presented in
Fig.~\ref{fig:dcr}. However, many more channels are open at 2.65~GeV
and it is not surprising that there is then a discrepancy of up to a
factor of three.

Data collected on a deuterium target are also presented in
Fig.~\ref{fig:dcr}. In this case the predictions of the three-channel
model for $K^+$ production in $pp$ interaction have been smeared over
the Fermi momentum, using the deuteron wave function derived from the
Bonn potential~\cite{BONN}. The Glauber shadowing effect, which was
discussed in detail for the $pd\to \eta X$ case, is included by
scaling the predictions for the sum of the cross sections on the
proton and neutron by a factor of 0.95~\cite{CHI1994}. Any
uncertainty in the size of this factor is small compared to the
experimental errors. Note that the other effect discussed there, of
$\eta$ conversion on the second nucleon, has no parallel for $K^+$
production.
%
%

\section{Discussion}
\label{discussion}%
\begin{figure}[hbt]\centering
\includegraphics[width=.9\columnwidth,clip=]{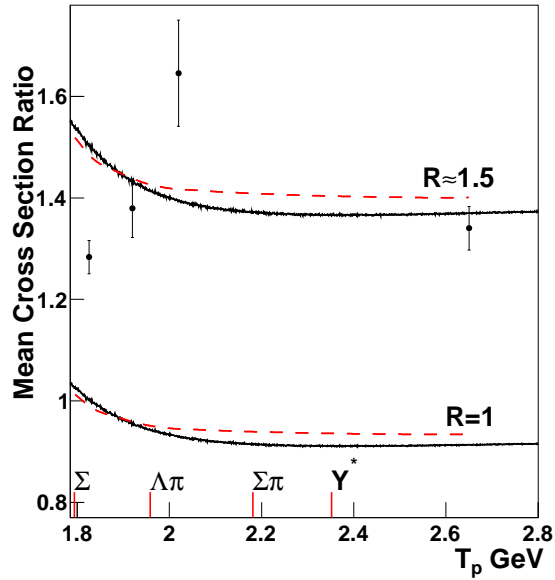}
\caption{\label{fig:money} (Color online) Ratio of $K^+$ production
cross sections on deuterium to hydrogen, averaged over the kaon
momentum range shown in Fig.~\ref{fig:dcr}. Errors due to the overall
normalizations are not included in the error bars. The black solid
line marked $R=1$ represents the ratio of Fermi-smeared $pp\to
K^+p\Lambda$ plus $pp\to K^+N\Sigma$ total cross sections divided by
the free hydrogen data. The red dashed line represents the same ratio
but evaluated from the differential spectra. The $R=1.5$ lines are
simply 1.5 times these values. The parameterizations of
Ref.~\cite{SIB2006}, together with the newer $\Sigma^+$ production
data~\cite{VAL2007,VAL2010}, were used in these estimations. In
addition to the thresholds indicated, kaon-pair production becomes
possible at about 2.5~GeV.}
\end{figure}

The ratio $d\sigma^{K^+}_{pd}/d\sigma^{K^+}_{pp}$ averaged over the
$K^+$ momentum is shown in Fig.~\ref{fig:money} for the four beam
energies investigated. It can be seen from the thresholds that are
indicated for the production of different strange baryonic states
that, unlike for the three lower energies, at 2.65~GeV there are far
more open channels than we have considered in our model. The
predictions of the model for the ratio of the Fermi-smeared $pp\to
K^+p\Lambda$ plus the $pp\to K^+N\Sigma$ total cross sections divided
by the free hydrogen data are shown by the curve $R=1$. Scaling this
by a factor of 1.5 gives results that are in reasonable accord with
the data. Simulations were also performed for the ratio of
differential cross sections shown in Fig.~\ref{fig:dcr} over the
proton beam energy range $T_p=1.8-2.65$~GeV, where the dependence on
kaon momentum is quite weak. It is seen from Fig.~\ref{fig:money}
that it makes very little difference in the modeling whether one
estimates the ratios of differential or total cross sections. This
gives us confidence to conclude that, despite relatively small
acceptance of ANKE, the observed ratio of differential cross sections
allows us to extract information on the difference between $K^+$
production on the proton and neutron. This is especially true for the
lower three energies, where the three-channel model should be
effective.

%
%
%
%

The ratio results presented here clearly indicate that inclusive
$K^+$ production on the neutron is less than on the proton at all the
four energies investigated. The best agreement between experimental
data and our model is obtained if the ratio of the total cross
sections for $K^+$ production in the two targets has a value of
$\sigma^{K^+}_{pd}/\sigma^{K^+}_{pp}=1.4\pm0.2$ which, after taking
the shadowing into account, means that the ratio for $K^+$ production
in $pn$ and $pp$ collisions
\begin{equation}\label{result}
\sigma_{pn}^{K^+}/\sigma_{pp}^{K^+}=0.5\pm0.2\,.
\end{equation}%
Although the ratio seems to depend only weakly upon the beam energy,
it should be noted that many different hyperons can be produced at
2.65~GeV and this can change the physics significantly.

At low energies, $\Lambda$ production dominates the inclusive cross
section and extra constraints then follow from isospin invariance.
Even if the $I=0$ amplitude in the $pn\to K^+n\Lambda$ reaction
vanishes, there would still be the
contribution from the $I=1$ term, so that%
\begin{equation}\label{isospin}%
\sigma^{\text{tot}}(pn\to K^+n\Lambda)/\sigma^{\text{tot}}(pp\to
K^+p\Lambda)>1/4.
\end{equation}%
This inequality is only exact for the total cross sections where
there can be no interference between the $I=0$ and $I=1$ waves. It is
not rigorous when there is a cut on the $K^+$ momentum and especially
on its angle which could, \emph{in principle}, lead to some
destructive interference over the acceptance of the ANKE
spectrometer. If this possibility is neglected and only $\Lambda$
production considered, Eq.~(\ref{result}) implies that
\begin{equation}\label{ratio2}
\sigma^{I=0}(NN\to KN\Lambda)/\sigma^{I=1}(NN\to KN\Lambda) = 1.0\pm0.8\,.
\end{equation}%
The difference between this and the result of Eq.~(\ref{result})
arises from the fact that, for a neutron target, half of the signal
would be associated with $K^0$ production.

The isospin ratio is much less than that determined for $\eta$
production, where a value of about twelve has been reported near
threshold~\cite{CAL1997}. It should therefore provide valuable
guidance for the modeling of kaon production in nucleon-nucleon
collisions as well as for experiments involving nuclear beams and/or
targets.

%
%

\section{Conclusions}
\label{conclusions}%
The double-differential cross sections for $K^+$ production have been
measured at four different proton beam energies using hydrogen and
deuterium targets. A value of
$d\sigma^{K^+}_{pd}/d\sigma^{K^+}_{pp}\approx1.4$ has been extracted
for the cross section ratio. This result is not incompatible with the
$\sigma(d\,\textrm{NaF} \to KX)/\sigma(p\,\textrm{NaF} \to KX)
=1.3\pm0.2$ obtained at 2.1~GeV per nucleon, though with the rather
complicated NaF target~\cite{SCH1989}. In our case the ratio seems to
depend weakly upon the proton beam energy and an average value of
$\sigma_{pn}^{K^+}/\sigma_{pp}^{K^+}=0.5\pm0.2$ was extracted from
the experimental data using a three-channel model. Taking into
account the unobserved $K^0$ production, this shows that strangeness
production near threshold is rather similar for the two isospin
channels of the initial nucleon-nucleon system.

To reduce the uncertainty in the cross section ratio it would be
necessary to measure the spectator proton $p_{\rm sp}$ in the $pd\to
p_{\rm sp} K^+n\Lambda$ reaction to ensure that the production had
taken place on the neutron. This is possible at ANKE through the use
of the silicon tracking telescopes and results will eventually be
available on inclusive $K^+$ production in the low energy
region~\cite{DZY2010}. Alternatively, the detection of all the fast
final particles in the $pd\to p_{\rm sp} K^0p\Lambda$ reaction would
allow the kinematics of spectator proton to be
reconstructed~\cite{KRA2009}.

%
%
\begin{acknowledgements}
We wish to thank the COSY machine crew and other members of the ANKE
collaboration for their help during the preparation and running of
the experiments reported here. This work has been partially supported
by BMBF, DFG(436 RUS 113/768), Russian Academy of Science, and COSY
FFE.
\end{acknowledgements}

%
%


\begin{thebibliography}{99}
%
\bibitem{BUS2004} M.~B\"{u}scher \textit{et al.}, Eur.\ Phys.\
    J.\ A \textbf{22}, 301 (2004).
%
\bibitem{LOP2007} X.~Lopez \textit{et al.}, Phys.\ Rev.\ C
    \textbf{75}, 011901(R) (2007).
%
\bibitem{RUD2005} Z.~Rudy \textit{et al.}, Eur.\ Phys.\ J.\ A
    \textbf{23}, 379 (2005).
%
\bibitem{GAU2010} S.~Gautam \textit{et al.}, J.\ Phys.\ G
    \textbf{37}, 085102 (2010).
%
\bibitem{TSU2000} K.~Tsushima, A.~Sibirtsev, A.~W.~Thomas and
    G.~Q.~Li, Phys.\ Rev.\ C \textbf{59}, 369 (1999).
%
\bibitem{HAR2005} C.~Hartnack, nucl-th/0507002 (2005).
%
\bibitem{BAL1996} J.~T.~Balewski \textit{et al.}, Phys.\ Lett.\ B
    \textbf{388}, 859 (1996); Phys.\ Lett.\ B \textbf{420}, 211
    (1998); Eur.\ Phys.\ J.\ A \textbf{2}, 99 (1998).
%
\bibitem{BIL1998} R.~Bilger \textit{et al.}, Phys. Lett. B
    \textbf{420}, 217 (1998).
%
\bibitem{SEW1999} S.~Sewerin \textit{et al.}, Phys.\ Rev.\
    Lett.\ \textbf{83}, 682 (1999).
%
\bibitem{KOW2004} P.~Kowina \textit{et al.}, Eur.\ Phys.\ J.\ A
    \textbf{22}, 293 (2004).
%
\bibitem{ABD2006} M.~Abdel-Bary \textit{et al.}, Phys.\ Lett.\
    B \textbf{632}, 27 (2006);
    M.~Abdel-Bary \textit{et al.}, Phys.\ Lett.\
    B \textbf{688}, 142 (2010).
%
\bibitem{LOU1961} R.~I.~Loutitt \textit{et al.}, Phys.\ Rev.\
    \textbf{123}, 1465 (1961).
%
\bibitem{BAL1988} A.~Baldini, V.~Flamino, W.~G.~Moorhead, and
    D.~R.~O.~Morison, Landolt-B\"{o}rnstein, New Series, Ed.\
    H.~Schopper (Springer-Verlag, Berlin, 1988).
%
\bibitem{VAL2007} Yu.~Valdau \textit{et al.}, Phys.\ Lett.\ B
    \textbf{652}, 27 (2007).
%
\bibitem{VAL2010} Yu.~Valdau \textit{et al.}, Phys.\ Rev.\ C
    \textbf{81}, 045208 (2010).
%
\bibitem{ABD2010a} M.~Abdel-Bary \textit{et al.}, Eur.\ Phys.\
    J.\ A \textbf{46}, 27 (2010).
%
\bibitem{SIB2006} A.~Sibirtsev, J.~Haidenbauer, H.-W.~Hammer
    and S.~Krewald, Eur.\ Phys.\ J.\ A \textbf{27}, 269 (2006).
%
\bibitem{ROZ2006} T.~Ro\.zek \textit{et al.}, Phys.\ Lett.\ B
    \textbf{643}, 251 (2006).
%
\bibitem{ABD2004} M.~Abdel-Bary \textit{et al.}, Phys.\ Lett.\ B
    \textbf{595}, 127 (2004).
%
\bibitem{ABD2007} M.~Abdel-Bary \textit{et al.}, Phys.\ Lett.\ B
    \textbf{649}, 252 (2007).
%
\bibitem{ANS1973} R.~E.~Ansorge, J.~A.~Charlesworth,
    N.~Intizar, W.~W.~Neale, and J.~G.~Rushbrooke, Nucl.\ Phys.\
    B \textbf{60}, 157 (1973);
    R.~E.~Ansorge, J.~R.~Carter, J.~A.~Charlesworth, W.~W.~Neale,
    and J.~G.~Rushbrooke, Phys.\ Rev.\ D \textbf{10}, 32 (1974).
%
\bibitem{Sigmaminus} E.~Shikov, Diploma thesis, St.\
Petersburg University (2009).
%
\bibitem{DZY2010} A.~Dzyuba \textit{et al.}, COSY proposal \#203 (2010).
%
\bibitem{KRA2009} M.~Krapp \textit{et al.}, IKP-J\"ulich Annual Report (2009).
%
\bibitem{BER1958} D.~Berkley and G.~B.~Collins, Phys.\ Rev.\
    \textbf{112}, 614 (1958).
%
\bibitem{SCH1989} S.~Schnetzer \textit{et al.}, Phys.\ Rev.\ C
    \textbf{40}, 640 (1989).
%
\bibitem{DEB1996} M.~Debowski \textit{et al.}, Z. Phys.\ A
    \textbf{356}, 313 (1996).
%
\bibitem{ABD2010b} S.~Abd El-Samad \textit{et al.}, Phys.\
    Lett.\ B \textbf{688}, 142 (2010).
%
\bibitem{BAL1999} F.~Balestra \textit{et al.}, Phys.\ Rev.\
    Lett.\ \textbf{83}, 1534 (1999).
%
\bibitem{FAL2005} G.~F\"aldt and C.~Wilkin, Eur.\ Phys.\ J.\ A
    \textbf{24}, 431 (2005).
%
\bibitem{IVA2006} A.~N.~Ivanov \textit{et al.},
    nucl-th/0509055.
%
\bibitem{MAI97} R.~Maier \textit{et al.}, Nucl.\ Instrum.\
    Methods Phys.\ Res.\ A \textbf{390}, 1 (1997).
%
\bibitem{KHO99} A.~Khoukaz \textit{et al.}, Eur.\  Phys.\ J.\
    D \textbf{5}, 275 (1999).
%
\bibitem{BAR01} S.~Barsov \textit{et al.}, Nucl.\ Instrum.\
    Methods Phys.\ Res.\ A \textbf{462}, 354 (2001).
%
\bibitem{BUS02} M.~B\"uscher \textit{et al.}, Nucl.\ Instrum.\
    Methods Phys.\ Res.\ A  \textbf{481}, 378 (2002).
%
\bibitem{DYM04} S.~Dymov \textit{et al.}, Part.\ Nucl.\ Lett.\
    \textbf{2}, 40 (2004).
%
\bibitem{UZI01} Yu.~Uzikov, ANKE note \#2 (2001).
%
\bibitem{YAS02} S.~Yaschenko \textit{et al.}, ANKE note \#7 (2002).
%
\bibitem{LEH04} I.~Lehmann \textit{et al.}, Nucl.\ Instrum.\
    Methods Phys.\ Res.\ A \textbf{530}, 275 (2004).
%
\bibitem{BAR04} S.~Barsov \textit{et al.}, Eur.\ Phys.\ A
    \textbf{21}, 521 (2004).
%
\bibitem{BONN} R.~Machleidt, K.~Holinde, and Ch.~Elster, Phys.\
    Rep.\ \textbf{149}, 1 (1987).
%
\bibitem{CHI1994} E.~Chiavassa \textit{et al.}, Phys.\ Lett.\ B \textbf{337},
192 (1994).
%
\bibitem{CAL1997} H.~Cal\'{e}n, \textit{et al.}, Phys.\ Rev.\
    Lett.\ \textbf{79}, 2642 (1997).
%
\end{thebibliography}
\end{document}